\documentclass[12pt]{iopart}

\usepackage{iopams}
\usepackage{graphicx,color}
\usepackage{color}

\newcommand{\be}{\begin{equation}}
\newcommand{\ee}{\end{equation}}

\begin{document}
\title{Extreme nonlocality with one photon}

 \author{Libby Heaney$^1$, Ad\'an Cabello$^2$,
Marcelo Fran\c ca Santos$^3$ and Vlatko Vedral$^{1,4}$}
\address{$^1$ Department of Physics, University of Oxford, Clarendon Laboratory, Oxford, OX1 3PU, UK}
\address{$^2$ Departamento de F\'{\i}sica Aplicada II, Universidad de Sevilla, E-41012 Sevilla, Spain}
\address{$^3$ Departamento de F\'isica, Universidade Federal de Minas Gerais, Belo Horizonte, Caixa Postal 702, 30123-970, MG, Brazil}
\address{$^4$ Centre for Quantum Technologies and Department of Physics, National University of
  Singapore, Singapore}
\ead{l.heaney1@physics.ox.ac.uk}
\ead{adan@us.es}


\begin{abstract}
Quantum nonlocality is typically assigned to systems of two or
more well separated particles, but nonlocality can also exist
in systems consisting of just a single particle, when one
considers the subsystems to be distant spatial field modes.
Single particle nonlocality has been confirmed experimentally
via a bipartite Bell inequality. In this paper, we
introduce an $N$-party Hardy-like proof of impossibility of
local elements of reality and a Bell inequality for local
realistic theories for a single particle superposed symmetrical
over $N$ spatial field modes (\emph{i.e.}, a $N$ qubit W
state). We show that, in the limit of large $N$, the Hardy-like
proof effectively becomes an all-versus nothing (or GHZ-like)
proof, and the quantum-classical gap of the Bell inequality
tends to be same of the one in a three-particle GHZ experiment.  We detail how to test the nonlocality in
realistic systems.
\end{abstract}

\maketitle


\section{Introduction}


Bell \cite{Bell:64}, and others \cite{Clauser:69}, constructed
a series of inequalities that are satisfied by pairs of
particles submitted to measurements $x$ (on
particle $A$) and $y$ (on particle $B$) with outcomes $a$ and
$b$, respectively, under the assumption that the joint
probability distributions can be written as
$P(x=a,y=b)=\sum_{\lambda}P(\lambda)P_A(x=a,\lambda)P_B(y=b,\lambda)$,
where $\lambda$ are pre-established classical correlations.
However, experiments \cite{Freedman:72, Aspect:82, Tittel:98,
Weihs:98, Rowe:01, Matsukevich:08} testing these inequalities
have consistently violated them, hence proving that it is
impossible to fully describe the world with theories satisfying
this assumption, called local realism.

Local realism is not the same as ``local elements of reality'',
defined as physical quantities whose outcomes can be predicted
with certainty from outcomes of space-like separated
measurements \cite{Einstein:35}. Both concepts are related, but
the former has the advantage of being independent of
predictions with certainty only existing in ideal experiments.
Bell's original target was proving the incompatibility between
local elements of reality and quantum mechanics, but the
violation of Bell's inequalities actually proves much more: it
proves that the world is incompatible with local realism.

The Greenberger-Horne-Zeilinger (GHZ) proof \cite{Greenberger:89}
is a proof of no local elements of reality, which can be
converted into a proof of no local realism through the
violation of a Bell inequality \cite{Mermin:90}. This violation
has been observed in the laboratory \cite{Pan:00, Zhao:03}. In
the case of the GHZ proof, the corresponding Bell inequality is
maximally violated by the GHZ class of many qubit entangled
states. A striking fact is that the degree of violation
increases exponentially with the number of particles.


Until relatively recently, nonlocality (\emph{i.e.}, the
impossibility of local realism) was presumed to be a property
of two (or more) well separated particles. However, Tan, Walls
and Collett \cite{Tan:91} pointed out that nonlocality could,
in principle, be determined via a Bell inequality test with a
single photon in a superposition of two distinct spatial field
modes. The fact that nonlocality could be considered an
intrinsic property of a single excitation of a quantum field
caused a flurry of discussions \cite{Hardy:94, Greenberger:95,
Dunningham:07, Cooper:08}, the upshot of which was the
experimental verification of entanglement of a single photon in
two separate sites \cite{Hessmo:04}.

Here we are concerned with the nonlocality generated by a
single particle as it is symmetrically superposed over an
increasing number of distant field modes to form the single
particle implementation of an $N$-qubit W state \cite{Dur:00}.
First, we prove the impossibility of local elements of reality
and then we derive a Bell inequality to experimentally detect
nonlocality.


In the proof of no local elements of reality, we
see that for a small number of sites (around $N\leq 10$) the
conflict with local elements of reality only occurs for a
fraction of events. However, as the number of sites tends to
infinity (or effectively $N>>10$), the conflict tends to occur for all events, in a
similar way it happens in a GHZ proof \cite{Pan:00}. This is
particularly surprising as the GHZ class is usually thought to
be the only capable of exhibiting this type of nonlocality.
More interestingly, while it is impossible to create a GHZ
state using less than three \emph{particles}, here we show that
a similar proof of no local elements of reality) can be
obtained for a state \emph{with just one particle}.

We start, in section (\ref{sec:theoproof}) with a
theoretical proof of no local elements of reality, then in
section (\ref{sec:theinequality}) we derive a Bell inequality
to experimentally test nonlocality, and explain how one may
check for the nonlocality of a single photon over $N$ sites in
practice. In order to give full weight to our results, in
section (\ref{sec:expimp}) we suggest how to implement the
required measurements in realistic conditions. Since we have
written this manuscript, we note that two other papers have
appeared that consider the nonlocality of a W state using our
results as a starting point \cite{Paraoanu:10, Ghirardi:11},
all the results are consistent with our original findings.


\section{Proof of no elements of reality for a single photon W state}
 \label{sec:theoproof}


We consider a system containing a single excitation, in this
case a photon, which is symmetrically superposed over $N$
sites. Each site represents a spatial field mode and we count
the number of photons in each mode \cite{Peres:95}. The state
of the system is then
\begin{equation}
\label{Eq:w-state}
|\psi_W\rangle_N=\frac{1}{\sqrt{N}}(|100\cdots0\rangle+|010\cdots0\rangle+\ldots+|
000\cdots1\rangle),
\end{equation}
where $|100\cdots0\rangle$ denotes that the photon occupies the
first site whilst the rest are empty. This is just the single
photon implementation of the $N$-qubit W state -- we have
chosen this implementation as it gives the most striking
violation of local realism that is known in a system containing
just one quanta.

Since only zero or one photons occupy each site at any
instance, the outcomes $z_i=\pm1$ of a Pauli, $\hat Z_i$,
measurement applied to the $i$th site, indicate whether the
photon is present in that site or not, \emph{i.e.}, $\hat Z_i
|0\rangle_i=|0 \rangle_i$ and $\hat Z_i |1\rangle_i=-|1\rangle_i$.
We will also consider local measurements in the Pauli $\hat X$
basis, whose outcomes $x_i=\pm1$ correspond to finding the
$i$th site in $|\pm\rangle =
\frac{1}{\sqrt{2}}(|0\rangle\pm|1\rangle)$ after the measurement.


\subsection{Local elements of reality of a W state}


In what follows we will use some properties of the W state
[points (i) and (ii) below] to derive a further measurement
setting [point (iii)] whose outcome is fixed for
models satisfying the criterion for local elements of reality
proposed by Einstein, Podolsky and Rosen \cite{Einstein:35}:
\emph{``If, without in any way disturbing a system, we can
predict with certainty (i.e., with probability equal to unity)
the value of a physical quantity, then there exists an element
of physical reality corresponding to this physical quantity''},
but is incompatible with some predictions of quantum
mechanics.


{\bf (i)} According to quantum mechanics, state
(\ref{Eq:w-state}) has the following property:
\begin{equation}
\label{eq:zs}
P_{\psi_W}( \underbrace{{z_i=\cdots=z_j}}_{N-1\;{\rm qubits}}=+1)=1,
\end{equation}
which is the probability of finding zero photons in $N-1$ sites
(although we cannot tell which ones) when we measure $\hat Z$
on all sites. It does not matter that we do not know which
$N-1$ sites will be empty until after the measurement (we will
discuss this point further shortly). Here the photon is found
in the $k$th site (which could be any of the $N$ sites: $k\in
1\dots N$), such that $P_W(z_k=-1)=1$: it is thus impossible for the
photon to be found on two or more sites simultaneously.

{\bf (ii)} We will define a further $N-1$ properties of state
(\ref{Eq:w-state}), but since all such properties are similar,
we will define first just one. The argument proceeds in a
counter-factual manner. In the measurement setting defined by
point (i) every site was measured in the local $\hat Z$ basis.
However, we could have instead measured two of the sites in the
Pauli $\hat X$ basis. We shall call these two sites the
$k$($\in1\dots N$)th site, which is the site where the particle
would have been found upon the particle number measurement, and
the $i$th site, which can be any other site. With this new
measurement setting, the outcomes, $x_k$ and $x_i$, are always
correlated, such that
\begin{equation}
\label{eq:zxs}
P_{\psi_W} ( {x_k=x_i} | \underbrace{{z_m=\cdots=z_r}}_{N-2\;{\rm
qubits}}=+1)=1.
\end{equation}
That is, the conditional probability of the local $\hat X_k$
and $\hat X_i$ measurements resulting in the same outcome,
given that the photon is not found in the remaining $N-2$
sites, is unity. Since we only have a single photon in the
state, there are always $N-2$ sites containing no photons, so
that we are conditioning on a property that is certain. For a
model satisfying the criterion of local elements of reality to
reproduce the W state $x_i=x_k$ must therefore hold.

The remaining $N-2$ properties have the same form as
(\ref{eq:zxs}), always with one $\hat X$ measurement on the
$k$th site, but with the other $\hat X$ measurement on a
different site each time:
\begin{equation}
\label{eq:alli}
P_{\psi_W} ( {x_j=x_k} | {z_m=\cdots=z_r}=+1)=1,\quad\quad\forall\, j\neq i,\, k.
\end{equation}
The fact that (\ref{eq:alli}) also holds follows, since we
could have instead measured the $j(\neq i)$th site along with
the $k$th site in the $\hat X$ basis to obtain $x_j=x_k$, and
so on. Using the properties (\ref{eq:zs}), (\ref{eq:zxs}) and
(\ref{eq:alli}) of the W state, we can conclude that the local
outcomes, $z$ and $x$, should have predefined values
(corresponding to local elements of reality) for all sites
before any measurement, since all of these statements occur
with certainty.

\begin{figure}[h] 
  \centering
  \includegraphics[width=4in]{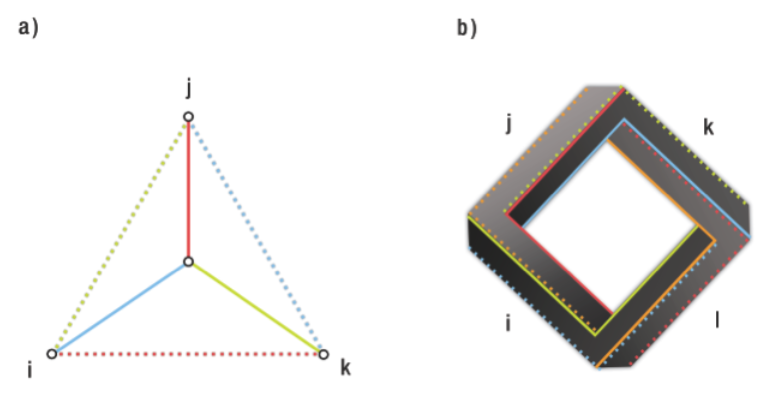}
  \caption{{\bf (a) Measurement outcomes inducing local elements of reality. }
The sites containing no photon upon a number measurement are labeled
$i$, $j$ and $k$ and situated at the edges of the triangle and the
site that would contain the photon is labeled $l$ and sits in the middle of the triangle.
A dotted line between two sites indicates that they both contain no photon.
Given this, it is guaranteed that the
remaining two sites are perfectly correlated in the $\hat X$ basis, which is represented by a solid line of the same colour linking these sites.
Models with local elements of reality always satisfy (\ref{fourLHV}), since all of the sites
are connected by solid lines. However, for four sites, quantum
mechanics violates this prediction half of the time.
 {\bf (b) Conflict with quantum mechanics. }Here each edge
represents a different site and the colours indicate different sets of photon number (dotted lines)
and superposition basis (solid lines) measurements.
Each measurement setting is itself consistent with local elements of reality. However, since the different observables
do not commute, one cannot simply add the settings and expect to obtain perfect correlations for
$X$ measurements on all of the sites.
The conflict with local elements of reality here is as transparent as the impossibility of this
Penrose square.
}
  \label{fig:logic}
\end{figure}

{\bf (iii)} We can now use the statements about local elements
of reality from points (i) and (ii) and the rules of classical
probability theory to construct a logical argument that any
theory satisfying the criterion of local elements of reality
must satisfy, namely:
\begin{equation}
P( \underbrace{{x_1=\cdots=x_N}}_{N\;{\rm qubits}} )=1,
\label{fourLHV}
\end{equation}
that an $\hat X$ measurement on all of the sites must result in
outcomes that are equal. This follows, since from point
(\ref{eq:zxs}) we can calculate
\begin{eqnarray}
P_{HV}(x_k=x_i) &=&P_{\psi_W}( {x_k=x_i} |\underbrace{{z_m=\cdots=z_r}}_{N-2\;{\rm
qubits}}=+1)P(z_m=\cdots=z_r=+1) \nonumber\\
 &+&P_{\psi_W}(x_k=x_i|z_m=-1\, z_n =\cdots=z_r=+1)\nonumber\\
 &&\quad\quad\quad\quad\times P(z_m=-1\, z_n =\cdots=z_r=+1)\nonumber\\
 &=& 1\times 1 + 0=1,
\end{eqnarray}
where $m\neq i\neq k$ and likewise for other pairs of sites.
The term $P_{\psi_W}(x_k=x_i|z_m=-1\, z_n
=\cdots=z_r=+1)P(z_m=-1\, z_n =\cdots=z_r=+1)$ is zero because
the photon can only be found in one site, which we have labeled
$k$, and thus cannot also be in site $m\neq k$. For a pictorial
depiction with four sites see Fig \ref{fig:logic} (a).


We now make a note about the use of counterfactuals in the
derivation of (\ref{fourLHV}). For simplicity, here we consider
just three sites, but our reasoning is straightforwardly
generalized to any number of sites. In point (i) we illustrated
the trivial property that upon a local particle number
measurement of all sites, the photon would only ever be found
in just one site; we labeled that site $k$, which could be
either site 1, 2 or 3, and there is a probability of one third
to be found in each site. We will now show why our argument
holds no matter which site the photon would have been found in.

In the model with local elements of reality of three sites, we
consider three rows of $\hat Z$ outcomes, namely:
\begin{eqnarray}
z_1 = -1\quad z_2=+1\quad z_3 =+1\nonumber\\
z_1=+1\quad z_2=-1\quad z_3 = +1\nonumber\\
z_1=+1 \quad z_2 = +1\quad z_3 = -1,
\end{eqnarray}
which are three different, mutually exclusive `worlds' that we
can work within to derive our prediction for these models. No
matter which world we find ourselves in when we make a  $\hat
Z$ measurement on all of the sites, the same
prediction [Eq. (\ref{fourLHV})] always follows. For instance,
if the photon would have been found in the second site,
$P(z_2=-1) = 1$, the properties $P(x_1=x_2|z_3=+1)=1$ and
$P(x_2=x_3|z_1=+1)=1$ lead us to conclude that
$P_{LER}(x_1=x_2=x_3)=1$. On the other hand, if the photon
is found in the first site, $P(z_1=-1)=1$
(\emph{i.e.}, the first row of values above), we would obtain
the same local realist prediction $P_{LER}(x_1=x_2=x_3)=1$ from
properties $P(x_1=x_2|z_3=+1)=1$ and $P(x_1=x_3|z_2=+1)=1$.
This is why we remarked in point (i) that it does not matter
which site the photon would have been found in after a $\hat Z$
measurement, as all the three `worlds' lead to the same
prediction.


\subsection{Quantum mechanical result}


Quantum mechanics can, however, contradict the prediction given
by Eq. (\ref{fourLHV}). The conflict for a three-qubit W state
was studied before \cite{Cabello:02} and the outcomes of local
$\hat X$ measurements on each of the sites was shown to
disagree with the outcome for models with local elements of
reality one quarter of the time. In this paper, we are
concerned with how the probability of violating the local
realist prediction Eq. (\ref{fourLHV}) scales with the number
of sites. Remarkably, as the single photon is superposed over
an increasing number of sites, even though the average number
of photons per site goes to zero, the probability of having a
conflict exponentially approaches unity:
\begin{eqnarray}
\label{eq:result}
P_v^{(N)} &=& 1-P_{\psi_W}(x_1=\cdots=x_N) \nonumber\\
&=&1 - \frac{N}{2^{N-1}},
\end{eqnarray}
(for instance, for twenty sites the probability to have a
conflict with local elements of reality is $P_v^{(N=20)}=
0.999962\ldots$). Thus, the outcomes for local $\hat X$
measurements on the state, $|\psi_W\rangle_N$, can {\it never}
be completed by elements of reality for sufficiently large $N$.

For a few sites (less than ten), ours is a Hardy-type proof
\cite{Hardy:93}. However, in the limit of {\it many} sites, the
W state created from a single photon behaves similarly to a GHZ
state and surprisingly demonstrates, for the first time to our
knowledge, an always--always--\ldots--always--never
contradiction. The ``always" clauses refer to the fact that for
any zero photon measurement of $N-2$ sites, the remaining two
sites will always be correlated in the $\hat X$ basis. The
``never" clause, on the other hand, implies that the
measurement of all sites in the $\hat X$ basis will never
result in all outcomes being the same, in the large $N$ limit
(see \cite{Cabello:02} for a more detailed explanation of this
notation). For all practical purposes our test is as
non-statistical as the GHZ test, since no matter how good the
measurement system is in the GHZ case, it will always have
probabilities of success that are below for e.g. $P_v^{(N=20)}=
0.999962$.

Especially surprising is the fact that such contradiction is
obtained using the properties of a non-stabilizing state
(\emph{i.e.}, a state without perfect correlations).  
Moreover,
we emphasise that our result is true for any $N$-qubit W state,
no matter how it is physically represented.  Note, that it has been shown \cite{Sen:03} that W states
comprised of $N>10$ qubits lead to a more robust (against noise
admixture) violations of local realism than the GHZ states, indicating further that large W states have very different properties to small W states. More investigations into the properties of large versus small W states would be fruitful.


\section{Bell inequality for a single photon W
state}
 \label{sec:theinequality}


To test this kind of nonlocality in actual experiments where
the \emph{perfect} correlations required to define local
elements of reality are never achieved, we have to derive the
$N$-party Bell inequality corresponding to the previous proof
(of no elements of reality). This Bell inequality holds for any
local theory (as defined in the introduction) and does not
require the notion of local elements of reality.

For $N=3$, the method in \cite{Cabello:10} show that, for any
local realistic theory,
\begin{eqnarray}
 \label{Winequality}
 \beta=&P(z_1=+1,z_2=+1,z_3=-1)+P(z_1=+1,z_2=-1,z_3=+1)\nonumber \\
 &+P(z_1=-1,z_2=+1,z_3=+1)\\
&-P(z_1=+1,x_2=+1,x_3=-1)-P(z_1=+1,x_2=-1,x_3=+1) \nonumber \\
&-P(x_1=+1,z_2=+1,x_3=-1)-P(x_1=-1,z_2=+1,x_3=+1) \nonumber \\
&-P(x_1=+1,x_2=-1,z_3=+1)-P(x_1=-1,x_2=+1,z_3=+1) \nonumber \\
&-P(x_1=+1,x_2=+1,x_3=+1)-P(x_1=-1,x_2=-1,x_3=-1) \le 0\nonumber,
\end{eqnarray}
The probabilities appearing in inequality (\ref{Winequality})
are exactly those involved in the argument of impossibility of
local elements of reality. The Bell inequality is tight: there
are local models which saturate the bound. For example, the one
in which $z_1=z_2=x_1=x_2=x_3=+1$ and $z_3=-1$.

The single photon W state (\ref{eq:alli}) violates inequality
(\ref{Winequality}). Specifically, it gives
\begin{equation}
\beta_{\psi_W}=\frac{1}{4},
\end{equation}
since
\begin{eqnarray}
 &P_{\psi_W}(z_1=+1,z_2=+1,z_3=-1)=P_{\psi_W}(z_1=+1,z_2=-1,z_3=+1)\nonumber \\
 &=P_{\psi_W}(z_1=-1,z_2=+1,z_3=+1)=\frac{1}{3},\nonumber
 \\
 &P_{\psi_W}(z_1=+1,x_2=+1,x_3=-1)=P_{\psi_W}(z_1=+1,x_2=-1,x_3=+1)=0,\nonumber
 \\
 &P_{\psi_W}(x_1=+1,z_2=+1,x_3=-1)=P_{\psi_W}(x_1=-1,z_2=+1,x_3=+1)=0,\nonumber
 \\
 &P_{\psi_W}(x_1=+1,x_2=-1,z_3=+1)=P_{\psi_W}(x_1=-1,x_2=+1,z_3=+1)=0,\nonumber
 \\
 &P_{\psi_W}(x_1=+1,x_2=+1,x_3=+1)=P_{\psi_W}(x_1=-1,x_2=-1,x_3=-1)=\frac{3}{8}.\nonumber
\end{eqnarray}


This inequality can be generalized to any $N>3$ as follows:
\begin{eqnarray}
 \label{WinequalityN}
 \Omega=&P(z_1=+1, \ldots, z_{N-1}=+1, z_N=-1) + \cdots + \nonumber\\
 &P(z_1=-1, z_2=+1, \ldots, z_N=+1) \nonumber \\
 &-P(z_1=+1, \ldots, z_{N-2}=+1, x_{N-1}=+1, x_N=-1) \nonumber \\
 &-P(z_1=+1, \ldots, z_{N-2}=+1, x_{N-1}=-1, x_N=+1)- \cdots \nonumber \\
 &-P(x_1=+1,x_2=-1,z_3=+1, \ldots, z_N=+1) \nonumber \\
 &-P(x_1=-1,x_2=+1,z_3=+1, \ldots, z_N=+1) \nonumber \\
 &-P(x_1=+1, \ldots, x_N=+1)-P(x_1=-1, \ldots, x_N=-1) \le 0.
\end{eqnarray}
The state (\ref{eq:alli}) violates inequality
(\ref{WinequalityN}). Specifically,
\begin{equation}
 \Omega_{\psi_W}=1 - \frac{N}{2^{N-1}},
\end{equation}
which tends to one when $N$ tends to infinity. A unity
quantum-classical gap is characteristic of the violation of the
three-party Bell inequality \cite{Mermin:90} (in terms of
probabilities of elementary propositions \cite{Cabello:10}) by
a GHZ state: there is a local model which mimics all but one of
the quantum predictions, but the price for the existence of
such model is that it gives a prediction, $P_{L}(x_1=+1,
\ldots, x_N=+1)+P_{L}(x_1=-1, \ldots, x_N=-1) =1$, which is the
opposite to the quantum one, $P_{\psi_W}(x_1=+1, \ldots,
x_N=+1)+P_{\psi_W}(x_1=-1, \ldots, x_N=-1) \rightarrow 0$ as
$N$ increases. This shows that, in the limit of large $N$, the
violation of the Bell inequality for the single photon W state
resembles the violation of the Bell inequality \cite{Mermin:90}
by a three-qubit GHZ state.

To test the nonlocality of a single photon W state in an
experiment, one needs to observe the violation of inequality
(\ref{WinequalityN}). The test goes as follows: one first tests
whether one has a single photon in the system by checking the
probabilities $P(z_1=+1, \ldots, z_{N-1}=+1, z_N=-1), \ldots,
P(z_1=-1, z_2=+1, \ldots, z_N=+1)$. Then, one tests the
probabilities $P(z_1=+1, \ldots, z_{N-2}=+1, x_{N-1}=+1,
x_N=-1), \ldots, P(x_1=-1,x_2=+1,z_3=+1, \ldots, z_N=+1)$.
Finally, one tests the probabilities $P(x_1=+1, \ldots,
x_N=+1)$ and $P(x_1=-1, \ldots, x_N=-1)$. To ensure locality,
these measurements should ideally be performed at a speed
faster than any communication between the sites.


\begin{figure}[t] 
 \centering
  \includegraphics[width=2.5in]{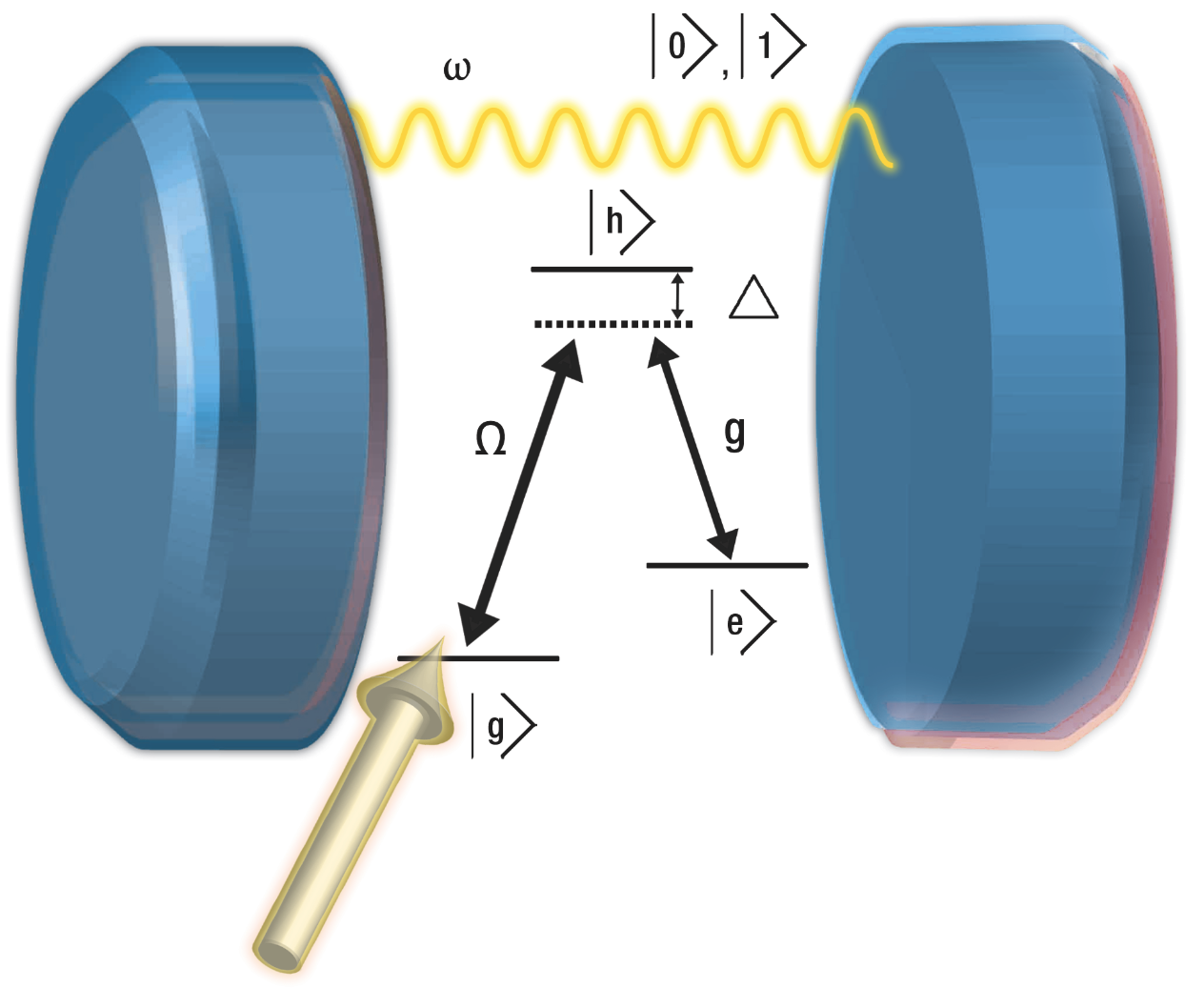}
\caption{Hadamard implementing unit: the combination of a three-level atom and an external
classical field allows for the Hadamard rotation in the Fock basis $|0\rangle, |1\rangle$
of the state of the cavity mode. The same atom can then be used to measure the state of
the cavity field. When $\Delta \sim 10\,g \sim 10\,\Omega$, level $|h\rangle$ can be
adiabatically eliminated and the off-resonant Raman transition between $|g\rangle$ and
$|e\rangle$ will strongly depend on the number of photons of the cavity mode \cite{Santos:01}.
The transition can then be tuned to resonance for a given Fock subspace, in our case
$|0\rangle,|1\rangle$, and the circuit described in \cite{Santos:05} can be applied
in order to implement deterministic rotations in this subspace as required for the
measurement in the $\hat{X}$ basis.}
 \label{fig:3}
\end{figure}


\section{Experimental implementations}
 \label{sec:expimp}


To implement our test, we need to first prepare a photon in a
symmetric superposition of many distant sites and then make
$\hat X$ and $\hat Z$ measurements on each of the $N$ sites. We
will describe below how to prepare this state with specific
implementations, but here we note that they are, in general,
considerably easier to create than the GHZ states. Measuring
$\hat Z_i$ means detecting whether there is a one photon in
site $i$. A measurement in the $\hat X$ basis can be achieved
by applying a Hadamard gate on the site and then measuring in
the $\hat Z$ basis. Note that performing these gates on each
site ends up adding photons to the system. However, this
addition is local and does not change the nonlocality of the
system as a whole, which is solely due to the spread of the
original photon.

The basic element of any experimental test of our work
therefore has to be a unit that is capable of deterministically
performing a Hadamard gate, i.e. creating a superposition of Fock states out of the vacuum. A
scheme achieving this is described in \cite{Santos:05}. It
relies on the selective manipulation of a three-level atom with
an external classical field and the cavity mode, which is
briefly pictured in Fig. (\ref{fig:3}) (for details of the selective
interaction see \cite{Santos:01}). For particular choices of
detuning $\Delta$ and coupling constants $g$ and $\Omega$ one
can adiabatically eliminate the third level $|h\rangle$
obtaining an effective resonant coupling between the atom and
the cavity mode that depends on the number of photons of the
latter. In fact, this coupling can be tuned in such a way that
only one chosen subspace of the entire Fock state basis, in our
case $|0\rangle,|1\rangle$, is resonant with the $|g\rangle
\rightarrow |e\rangle$ atomic transition. Then, the quantum
circuit described in \cite{Santos:05} can be used to
deterministically rotate states $|0\rangle$ and $|1\rangle$
respectively into states $(|0\rangle + |1\rangle)/ \sqrt{2}$
and $(-|0\rangle + |1\rangle)/\sqrt{2}$.

Once one is able to implement a Hadamard rotation, this basic
element can be used in a number of different ways to perform
tests of nonlocality suggested in this paper. Here we briefly
describe an implementations, which is well
within the current experimental state of the art. 

It involves using an unbiased multiport
beam splitter \cite{Mattle:95} to create the W
state (here the aforementioned difficulty of preparing GHZ
states is apparent; a GHZ state would require no photons in any
port to coherently be superposed with one photon in every
port!).
Note that a heralded four mode single photon W state has
recently been created via a sequence of beamsplitters
\cite{Papp:09}, which would allow to test the Hardy-type regime
of our Bell theorem Eq. (\ref{eq:result}). To each port we
couple an optical fiber, which guides the photon to a cavity
\cite{Cirac:97, vanEnk:98} containing the above described unit
Hadamard element. In this way, we are able to perform both the
$\hat{X}$ and $\hat{Z}$ measurements in each cavity and
therefore test our violations of nonlocality. In order to
guarantee the existence of the photon entering the multiport
beam splitter, one can generate a twin beam in parametric down
conversion and use one of the photons as the trigger for the
experiment.

Another method, utilising unbalanced homodyne detection, has
recently been put forward to test our proposal
\cite{Laghaout:10}. Within this scheme, a quantum efficiency of
69\% is required for a single photon, three-mode W state to
exhibit a detectable violation of local realism. It would be
interesting to investigate further how this efficiency scales
with the number of sites.

A much more challenging experiment would be to test the W
nonlocality with one massive (instead of massless) particle.
There is a long standing lively debate \cite{Spekkens:07,
Heaney:09, Heaney:09a} about whether mode entanglement of massive particles can be used for tests of nonlocality, which is due to the presence of superselection rules that are not present in the photonic case.  We hope that our work stimulates further research
in this important direction.


\section{Conclusions}


In this paper, we have derived a Hardy-like proof of impossibility of local elements of reality for an N-site single photon W state.  We have shown that in the limit of a large number of sites this proof effectively becomes an all-versus-nothing proof, similar to the GHZ test of nonlocality.  We have derived a Bell inequality that allows to experimentally check for local realistic theories and we point out how this test could be implemented in realistic systems.


\section*{Acknowledgements}


L. H. acknowledges the financial support of EPSRC, UK. A. C.
acknowledges support from the Spanish MICIN Project
FIS2008-05596 and the Wenner-Gren Foundation. M. F. S. is
supported by the CNPq and Fapemig. V. V. is grateful for
funding from the National Research Foundation and the Ministry
of Education (Singapore). L. H. and V. V. thank the UFMG,
Brazil for its hospitality during the period when this work was
produced.


\section*{References}



\begin{thebibliography}{99}

\bibitem{Bell:64}
	Bell J 1964
	{\it Physics} {\bf 1} 195

\bibitem{Clauser:69}
	Clauser J F, Horne M A, Shimony A and Holt R A 1969
	{\it Phys. Rev. Lett.} {\bf 23} 880
 	
\bibitem{Freedman:72}
	Freedman S J and Clauser J F 1972
	{\it Phys. Rev. Lett.} {\bf 28} 938

\bibitem{Aspect:82}
	 Aspect A, Dalibard J and Roger G 1982
	 {\it Phys. Rev. Lett.} {\bf 49} 1804

\bibitem{Tittel:98}
	Tittel W, Brendel J, Zbinden H and Gisin N 1998
	 {\it Phys. Rev. Lett.} {\bf 81} 3563
	
\bibitem{Weihs:98}
	Weihs G, Jennewein T, Simon Ch, Weinfurter H and Zeilinger
A 1998
	{\it Phys. Rev. Lett.} {\bf 81} 5039
	
\bibitem{Rowe:01}
	 Rowe M A, {\it et al.} 2001	
	{\it Nature} {\bf 409} 791
	
\bibitem{Matsukevich:08}
	Matsukevich D N, Maunz P, Moehring D L, Olmschenk S and
Monroe C 2008
	{\it Phys. Rev. Lett.} {\bf 100} 150404

\bibitem{Einstein:35}
	Einstein A, Podolsky B and Rosen N 1935
	{\it Phys. Rev.} {\bf 47} 777

\bibitem{Greenberger:89}
	Greenberger D M, Horne M A and Zeilinger A 1989
    In
    {\em Bell's Theorem, Quantum Theory, and Conceptions of the Universe}
    (Kluwer Academic, Dordrecht), p~69


\bibitem{Mermin:90}
	Mermin N D 1990
	{\it Phys. Rev. Lett.} {\bf 65} 1838

\bibitem{Pan:00}
	Pan J W, {\it et al.} 2000
	{\it Nature} {\bf 403} 515
	
\bibitem{Zhao:03}
	 Zhao Z, Yang T, Chen Y-A,
	 Zhang A-N, \.{Z}ukowski M and Pan J W 2003
	{\it Phys. Rev. Lett.} {\bf 91} 180401
	
\bibitem{Tan:91}
	Tan S M, Walls D F and Collett M J 1991
	{\it Phys. Rev. Lett.} {\bf 66} 252
	
\bibitem{Hardy:94}
	Hardy L 1994
	{\it Phys. Rev. Lett.} {\bf 73} 2279

\bibitem{Greenberger:95}
	Greenberger D M, Horne M A and Zeilinger A 1995
	{\it Phys. Rev. Lett.} {\bf 75} 2064
	
\bibitem{Dunningham:07}
	Dunningham J A and Vedral V 2007
	{\it Phys. Rev. Lett.} {\bf 99} 180404
	
\bibitem{Cooper:08}
	Cooper J J and Dunningham J A 2008
	{\it N. J. Phys.} {\bf 10} 113024 	

\bibitem{Dur:00}
    D\"{u}r W, Vidal G, and Cirac J I 2000
    {\it Phys. Rev.~A} {\bf 62} 062314

\bibitem{Hessmo:04}
    Hessmo B, Usachev P, Heydari H and Bj{\"o}rk G 2004
    {\it Phys. Rev. Lett.} {\bf 92} 180401
 	
\bibitem{Paraoanu:10}
	Paraoanu G S 2010
	arXiv:1011.1795v1.	
	
\bibitem{Ghirardi:11}
	Ghirardi G C 2011
	arXiv:1101.5252v1.
	


\bibitem{Peres:95}
	Peres A 1995
	{\it Phys. Rev. Lett.} {\bf 74} 4571
 	
\bibitem{Cabello:02}
	Cabello A 2002
	{\it Phys. Rev. A} {\bf 65} 032108

\bibitem{Hardy:93}
	Hardy L 1993
	{\it Phys. Rev. Lett.} {\bf 71} 1665

\bibitem{Sen:03}
 	Sen(De) A, Sen U, Wie\`{s}niak M, Kaszlikowski D and \.{Z}ukowski M 2003
	{\it Phys. Rev. A} {\bf 68} 062306 	
	


\bibitem{Cabello:10}
    Cabello A, Severini A and Winter A 2010
    arXiv:1010.2163v2.

\bibitem{Santos:05}
	Santos M F 2005
 	{\it Phys. Rev. Lett.} {\bf 95} 010504
	
\bibitem{Santos:01}
	Santos M F, Solano E and de Matos Filho 2001
	{\it Phys. Rev. Lett.} {\bf 87} 093601
	

\bibitem{Papp:09}
	Papp S B, Choi K S, Deng H, Lougovski P,  van Enk S J and
    Kimble H J 2009
	{\it Science} {\bf 324} 764
	
\bibitem{Mattle:95}
	Mattle K, Michler M, Weinfurter H, Zeilinger A and Zukowski M 1995
	{\it Appl. Phys. B: Laser Opt} {\bf 60} S111 (1995).
			
\bibitem{Cirac:97}
	Cirac J I, Zoller P, Kimble H J and Mabuchi H 1997
	{\it Phys. Rev. Lett.} {\bf 78} 3221

\bibitem{vanEnk:98}
	van Enk S J, Cirac J I and Zoller P 1998
	{\it Science} {\bf 279} 205


\bibitem{Laghaout:10}
    Laghaout A and Bj{\"o}rk G 2010
    {\it Phys. Rev. A} {\bf 81}, 033823
	
\bibitem{Spekkens:07}
    Bartlett S D, Rudolph T and Spekkens R W 2007
	{\it Rev. Mod.  Phys.}  {\bf 79} 555 	

\bibitem{Heaney:09}	
	Heaney L and Anders J 2009
	{\it Phys. Rev. A} {\bf 80} 032104 	

\bibitem{Heaney:09a}	
	Heaney L and Vedral V 2009
	{\it Phys. Rev. Lett.} {\bf 103} 200502
	

\end{thebibliography}
\end{document}